\documentstyle[prd,aps]{revtex}

\begin{document}

\title{Resonant Spin-Flavor Conversion of Supernova Neutrinos \\
and \\ Deformation of the Electron Antineutrino Spectrum \\}

\vspace{1cm}

\author{Tomonori Totani$^1$ and Katsuhiko Sato$^{1,2}$}
\address{$^1$Department of Physics, School of Science,
The University of Tokyo, \\
7-3-1 Hongo, Tokyo 113, Japan. \\
e-mail: totani@utaphp1.phys.s.u-tokyo.ac.jp \\
$^2$Research Center for the Early Universe, School of Science,
The University of Tokyo, \\
7-3-1 Hongo, Tokyo 113, Japan.}

\maketitle

\vspace{-7cm}
\begin{flushright}
UTAP-233/96 \\
RESCEU-15/96
\end{flushright}

\vspace{7cm}
\begin{center}
\bf Abstract
\end{center}

\begin{abstract}
The neutrino spin-flavor conversion of $\bar\nu_e \leftrightarrow
\nu_\mu$ which is induced by
the interaction of the neutrino magnetic moment and 
magnetic fields in the collapse-driven supernova is investigated in
detail.  We calculate the conversion probability
by using the latest precollapse models of Woosley and Weaver (1995),
and also those of Nomono and Hashimoto (1988),
changing the stellar mass and metallicity in order to 
estimate the effect of the astrophysical uncertainties.
Contour maps of the conversion probability
are given for all the models as a function of 
neutrino mass squared difference over neutrino energy 
($\Delta m^2 / E_\nu$)
and the neutrino magnetic moment times magnetic fields ($\mu_\nu B$). 
The expected deformation of 
the $\bar\nu_e$ spectrum can be clearly seen from these maps,
and some qualitative features which will be useful in the 
future observation are summarized.
It is shown that in the solar metallicity models
some observational effects are expected 
with $\Delta m^2 = 10^{-5}$--$10^{-1}$ [eV$^2$]
and $\mu_\nu \agt 10^{-12} (10^9 {\rm G} / B_0)$ [$\mu_B$],
where $B_0$ is the strength of the magnetic fields
at the surface of the iron core, and $B_0 = 10^9$ [Gauss] is a 
reasonable value which is inferred from the observed magnetic fields in
white dwarfs.
We also find that although the dependence on the stellar models or 
stellar mass is not so large, the metallicity of precollapse stars
has considerable
effects on this conversion. In lower metallicity stars,
strong precession between $\bar\nu_e$ and $\nu_\mu$ occurs
with small $\Delta m^2 / E_\nu$ ($\alt 10^{-8}$
[eV$^2$/MeV]), and conversion probability changes periodically with $B_0$.
Such effects may be seen in a supernova in the
Large or Small Magellanic Clouds, and should be taken into account
when one considers an upper bound on $\mu_\nu$ from the SN1987A data.
\end{abstract}


\newpage

\section{Introduction}
Copious neutrino emission from collapse-driven supernovae
attracts significant attention because it provides rich
information not only on the mechanism of supernovae but also
on the neutrino physics through a number of events captured
in some underground neutrino detectors, such as the Super-Kamiokande
(SK)\cite{SK}. The most noteworthy subject on the nature of neutrinos
is the mass of neutrinos
and oscillations between different flavors induced by the mass difference.
However, the ordinary matter oscillation (well known as the MSW
effect\cite{MSW}) has its effect only on neutrinos but not on 
antineutrinos under the direct mass hierarchy, and the vacuum 
oscillation is not observable unless the mixing angle is unnaturally 
large compared with that of the quark sector.
In this case, electron antineutrinos
($\bar\nu_e$'s), which is the most detectable in a water \v{C}erenkov
detector, do not undergo any oscillation.
One of some possibilities that $\bar\nu_e$'s would oscillate or
be converted into other species of neutrinos is the neutrino magnetic
moment. If the neutrinos have a nonvanishing magnetic moment, it
couples the left- and right-handed neutrinos, and interaction
with sufficiently strong magnetic fields induces the 
precession between neutrinos with different chiralities
in the inner region of the collapse-driven supernova
\cite{Cisneros,Fujikawa-Shrock}.
In general, non-diagonal elements of the magnetic moment matrix
are possible, and neutrinos can be changed into different flavors
by this flavor changing moment\cite{Schechter-Valle}.
Furthermore, with the additional 
effect of the coherent forward scattering by
dense matter in the collapsing star, 
neutrinos can be resonantly converted\footnote{The 
precession itself is suppressed by the matter potential.} into
neutrinos with different chiralities
\cite{Lim-Marciano,Akhmedov,Voloshin,Akhmedov-Berezhiani,Peltoniemi,%
Athar-etal} by the mechanism similar
to the MSW effect. This resonant
spin-flavor conversion induced by the neutrino magnetic moment
may drastically deform 
the spectrum of electron antineutrinos ($\bar\nu_e$'s)
in the water \v{C}erenkov detectors.
The earlier publications have shown that in the future experiment
this effect will be 
observable with inner magnetic fields of some reasonable strength,
if there is a magnetic 
moment a little smaller than the current astrophysical upper 
limits from the argument of the stellar cooling due to
the plasmon decay\footnote{This constraint refers to the norm of the 
neutrino magnetic moment matrix, 
$(\sum_{i,j}|\mu_{ij}|^2)^{1/2}$, i.e.,
this includes the flavor changing moment.}: 
$\mu_\nu \alt 10^{-11}$--$10^{-10} 
\mu_B$, where $\mu_B$ is the Bohr magneton \cite{Fukugita-Yazaki,%
Raffelt}. The magnetic moment
of neutrinos in the standard electro-weak theory with small neutrino
masses is very small due to 
the chirality suppression; for example, 
the standard SU(2)$_L \times$U(1) model with a singlet
right-handed neutrino gives $\mu_\nu \sim 3\times 10^{-19}
(m_\nu/1{\rm eV}) \mu_B$, far below the experimental/astrophysical 
upper bounds\cite{Fujikawa-Shrock,Marciano-Sanda,Lee-Shrock,Petkov}.
However, 
some particle-physics models\cite{Fukugita-Yanagida,Babu-Mathur} 
have been proposed
in order to give a large magnetic moment of $\sim 10^{-11} \mu_B$ 
which would explain\cite{VVO,Akhmedov-b,Inoue-Dron}
the anticorrelation
between the time variability of the solar neutrino flux and the sun spot
numbers suggested in the $^{37}$Cl experiment\cite{Davis}.
(The anticorrelation in the Cl experiment, however, 
has not yet been statistically settled.)
Therefore, the influence
of a large magnetic moment on various physical or astrophysical
phenomena including collapse-driven supernovae deserves more detailed
investigation. Discovery of a large magnetic moment
of neutrinos indicates that there exist interactions which
violate the chirality conservation beyond the standard theory.

In this paper, the resonant spin-flavor conversion between
right-handed
$\bar\nu_e$'s and left-handed mu or tau neutrinos ($\nu_\mu$ 
or $\nu_\tau$'s)
is studied assuming that the neutrino is the Majorana particle.
In general, the matter potential suppresses the interaction of
the magnetic moment and magnetic fields because of the generated
difference of the energy levels. However,
Athar et al.\cite{Athar-etal} pointed out that 
the resonant conversion of this mode ($\bar\nu_e \leftrightarrow \nu_\mu$)
occurs quite effectively in the region above the iron core and below the 
hydrogen envelope of collapsing stars, namely, 
in the O+Si, O+Ne+Mg, O+C, and He layers (hereafter 
referred to `the isotopically neutral region'). 
The reason is that the effective matter potential for the 
$\bar\nu_e \leftrightarrow \nu_\mu$ mode
is given in the form proportional to the value of ($1-2Y_e$),
where $Y_e$ is the electron number fraction per nucleon,
and $Y_e$ is very close to 0.5 in this region (typically,
$(1-2Y_e) \sim 10^{-4}$--$10^{-3}$);
the matter effect is therefore strongly suppressed compared with the 
magnetic interaction, and the adiabaticity condition
becomes considerably less stringent. 
Athar et al.\cite{Athar-etal} have shown
that assuming $\mu_\nu (=\mu_{\bar\nu_e \nu_\mu})
\sim 10^{-12} \mu_B$, this resonant conversion
would occur with some reasonable assumptions about magnetic fields 
in a star. We also consider $\mu_\nu$ around this value\footnote{%
There is a further stringent constraint on the transition magnetic
moment of massive neutrinos from observation of the 21-cm
(hyperfine) radiation from neutral hydrogen gas in external
galaxies: $\mu_\nu \leq
1.7 \times 10^{-15}$ for the neutrino masses above 30 eV 
\cite{Raphaeli-Szalay,Nusseinov-Raphaeli}. However, we consider
only the range of $\Delta m^2 \alt 1$ [eV$^2$] and this upper bound
does not constrain our analysis.}.

In order to judge the deformation of an observed $\bar\nu_e$ spectrum 
as the evidence of the existence of the neutrino magnetic 
moment, it is necessary that the conversion probability is 
calculated with high accuracy in a wide range of some parameters
such as the mass of neutrinos or the magnetic fields. However,
only rough estimates or demonstrations in some cases are given
in the earlier publications and the relation between the shape of
the deformed spectrum and the physical parameters has not yet been
clarified. Therefore we make the contour maps of 
the conversion probability of $\bar\nu_e \leftrightarrow \nu_\mu$
for some used models of precollapse stars as a function of 
the two parameters: $\Delta m^2 / E_\nu$ and $\mu_\nu B_0$,
where $\Delta m^2$ is the neutrino mass squared difference,
$E_\nu$ the neutrino energy, 
and $B_0$ the magnetic field at the surface of the iron core.
The expected observational effects can be clearly understood by these maps.
Some examples of spectral deformation
are also calculated and qualitative features which will
be useful for the future observation are summarized by using
these maps.

It is apparent that the deviation of the value of $Y_e$ from 0.5
in the isotopically neutral region
is quite important, and this value is strongly dependent
on the isotopic composition. Since almost all nuclei in the
isotopically neutral region are symmetric
in the number of neutrons and protons, 
this deviation is determined by rarely existent nuclei and the accurate
estimate of this deviation is quite difficult. Therefore,
the astrophysical uncertainty in $(1-2Y_e)$ should be discussed.
We use the latest 15 and 25 $M_\odot$ precollapse
models of Woosley and Weaver (hereafter WW) \cite{WW1995}
which include no less than 200 isotopes. Such a large number of isotopes
have never been used previously in the calculation of $(1-2Y_e)$. 
It is also expected that this value strongly depends on the stellar
metallicity, and hence we use the WW models 
with the two different metallicities: the solar and zero 
metallicity, and the metallicity effect is investigated.
Also the models of 4 and 8 $M_\odot$ helium core of Nomoto \&
Hashimoto \cite{NH1988} (hereafter NH)
are used, which correspond approximately
to 15 and 25 $M_\odot$ main sequence stars, and the model dependence
of $(1-2Y_e)$ is discussed. 

We consider $\Delta m^2$ smaller than about 1 [eV$^2$], therefore
the resonance occurs above the surface of the iron core.
The resonance in the iron core and its implications on the dynamics of 
the supernova considered in a recent preprint\cite{Akhmedov-prep}
are not discussed here.
The global structure of magnetic field is assumed to be a dipole moment,
and the strength of the magnetic field is normalized at 
the surface of the iron core with the values of $10^8$--$10^{10}$
[Gauss], which are inferred from the observation of the magnetic fields
on the surface of white dwarfs. 
Throughout this paper, we consider the conversion between
two generations for simplicity. Because $\nu_\mu$ and $\nu_\tau$ 
can be regarded as identical particles
in the collapse-driven supernova, our results also apply to the 
conversion of $\bar\nu_e \leftrightarrow \nu_\tau$.
Derivation of the equation which describes the propagation of 
neutrinos and evolution of conversion probability is given in
section \ref{sec:osc-mech}, 
and the profile of the effective matter potential and 
magnetic fields are given in section \ref{sec:astro} by using the
precollapse models of massive stars.  
Qualitative features of the conversion
are also discussed in this section. Numerical results are given in section 
\ref{sec:results},
and spectral deformation is also discussed.  Discussion and conclusions
are given in sections \ref{sec:discussion} and \ref{sec:summary},
respectively.

\section{Formulations}
\label{sec:osc-mech}
The interaction of the magnetic moment of neutrinos
and magnetic fields is described as 
\begin{equation}
<(\nu_i)_R|H_{\rm int}|(\nu_j)_L> = \mu_{ij} B_{\bot} \ ,
\end{equation}
where $\mu_{ij}$ is the magnetic moment matrix, $B_\bot$ the 
magnetic field transverse to the direction of propagation, 
$(\nu)_R$ and $(\nu)_L$ the right- and left-handed neutrinos,
respectively, and $i$ and 
$j$ denote the flavor eigenstate of neutrinos, i.e., e, $\mu$, and
$\tau$. The magnetic moment interacts only with transverse
magnetic fields.
If neutrinos are the Dirac particles, right-handed neutrinos
and left-handed antineutrinos do not interact
with matter and therefore undetectable. The 
conversion into these sterile neutrinos due to the magnetic moment
suffers strong constraints from the observation of neutrinos from 
SN1987A by the Kamiokande II \cite{SN1987A-Kam} and 
IMB \cite{SN1987A-IMB}, and also
from the argument on energy transportation in the collapse-driven
supernova\cite{Voloshin,Peltoniemi,Dar,Lattimer-Cooperstein,%
Barbieri-Mohapatra,Notzold}. 
On the other hand, if neutrinos are the Majorana 
particles, as assumed in this paper, 
$\nu_R$'s are antineutrinos and interact with
matter, and the constraint becomes considerably weak.
The diagonal magnetic moment is forbidden for the Majorana
neutrinos, and therefore only the conversion between
different flavors is possible, e.g., $(\bar\nu_e)_R
\leftrightarrow (\nu_{\mu, \tau})_L$. As mentioned in introduction,
we investigate this mode because 
the conversion of this mode occurs quite effectively in the 
isotopically neutral region and also $\bar\nu_e$'s
are most easily detected in the water \v{C}erenkov detectors.

In dense matter of the collapsing stars, the coherent
forward scattering by matter leads to the effective potential
for neutrinos, and this potential for each type of neutrinos
is determined according to the 
weak interaction theory.
The potential due to scattering with electrons is given as
(including both the charged- and neutral-current interactions)
\begin{equation}
V = \pm \sqrt{2} \ G_F \ (\pm \frac{1}{2} + 2 \sin^2 \theta_W) \ n_e \ ,
\end{equation}
where $n_e$ is the number density of electrons, 
$G_F$ the Fermi coupling constant, and $\theta_W$ the 
Weinberg angle. The $\pm$ sign in the parentheses refers to $\nu_e$ (+)
and $\nu_{\mu, \tau}$ ($-$), and that in front to $\nu$ (+) and 
$\bar\nu$ ($-$). In the ordinary flavor oscillation 
($\nu_e \leftrightarrow \nu_{\mu, \tau}$), the effective 
potential is only due to the charged-current scattering by
electrons because the effect of 
neutral-current interactions is the same for all flavors.
However, we have to consider the neutral-current interaction
in the conversion of $\nu$ and $\bar\nu$,
because of the opposite signs of the potential.
Therefore the neutral-current scattering by nucleons should also be
included, that is
\begin{equation}
V = \pm \sqrt{2} \ G_F \ (\frac{1}{2}-2 \sin^2 \theta_W) \ n_p
    \mp \sqrt{2} \ G_F \ \frac{1}{2} \ n_n \ ,
\end{equation}
where $n_p$ is the number density of protons, $n_n$ that
of neutrons. The $\pm$ or $\mp$ signs refer to $\nu$ (upper)
and $\bar\nu$ (lower) for all three flavors of neutrinos.
We do not have to consider the form factor of nuclei
because the relevant
interaction is forward scattering and there is no
momentum transfer. 
The isotopically neutral region is far beyond the neutrino sphere
and neutrinos go out freely in this region; hence 
we do not have to consider the neutrino-neutrino scattering.
By using the charge neutrality,
the difference of the potentials for $\bar\nu_e$'s
and $\nu_\mu$'s (or $\nu_\tau$'s)
which we are interested in is as follows:
\begin{equation}
\Delta V \equiv V_{\bar\nu_e} - V_{\nu_\mu} =   
\sqrt{2}G_F\rho/m_N (1 - 2 Y_e) \ ,
\end{equation}
where $\rho$ is the density, 
$m_N$ the mass of nucleons,
and $Y_e = n_p / (n_p + n_n)$.
Now the time evolution of the mixed state of $\bar\nu_e$
and $\nu_\mu$ is described
by the following Schr\"{o}dinger equation:
\begin{equation}
i\frac{d}{dr}\left( \begin{array}{c} \bar\nu_e \\ \nu_\mu
\end{array}\right) = \left(
\begin{array}{cc} 0     & \mu_\nu B_\bot \\
                  \mu_\nu B_\bot & \Delta H \end{array} \right)
\left( \begin{array}{c} \bar\nu_e \\ \nu_\mu 
\end{array}\right) \ ,
\label{eq:schrodinger}
\end{equation}
and $\Delta H$ is defined as:
\begin{equation}
\Delta H \equiv \frac{\Delta m^2}{2 E_\nu} \cos 2 \theta 
- \Delta V \ , 
\end{equation}
where $E_\nu$ is the energy of neutrinos, $\Delta m^2 =
m^2_{2} - m^2_{1}$, $\theta$ the angle of the vacuum generation
mixing, and $r$ the radius from the center of the star.
Here we consider only $\bar\nu_e$ and $\nu_\mu$, but this equation
is actually a truncation of the original 4-component ($\nu_e$,
$\nu_\mu$, and antineutrinos) equation (see ref. \cite{Lim-Marciano}).
The neutrino masses, $m_1$ and $m_2$ are
those in the mass eigenstates ($m_2 > m_1$).
The direct mass hierarchy is assumed here and 
therefore $\Delta m^2$ is positive. The other terms have
their standard meanings and the units of $ c = \hbar = 1 $  are used.
Also note that we can subtract an arbitrary constant times the unit
matrix from the Hamiltonian, which does not affect the probability
amplitudes.
In the MSW flavor oscillation, there appears
the term of generation mixing, $\Delta m^2 \sin 2 \theta
/ 4E_\nu$, in the off-diagonal elements of the Hamiltonian;
however, this term does not appear in this spin-flavor conversion
between neutrinos and antineutrinos.
In the following, 
$\mu_\nu$ and $\cos 2 \theta$ are
set to be $10^{-12}\mu_B$ and 1, respectively, and 
the scaling of $B$ or $\Delta m^2$
with respect to other values of $\mu_\nu$ or $\cos 2 \theta$
is obvious.
The resonant spin-flavor conversion occurs when the difference of
the diagonal elements
in the Hamiltonian vanishes, and hence the resonance condition is 
given as $\Delta H = 0$.
By using this equation, the probability of conversion can be
calculated provided that $\rho(r), Y_e(r)$, and $B_\bot(r)$ are known.

\section{Astrophysical Aspects}
\label{sec:astro}
\subsection{Effective Matter Potential}
In this section, we consider the effective matter potential 
in the isotopically neutral region.
The value of $(1-2Y_e)$ which we are interested in is easily
calculated as:
\begin{equation}
Y_e - \frac{1}{2} = \sum_i \left( \frac{Z_i}{A_i} - \frac{1}{2} \right) X_i
\ ,
\end{equation}
where $Z_i$, $A_i$, and
$X_i$ are the atomic number, mass number, and
the mass fraction of the $i$-th isotope, respectively, and 
the subscript $i$ runs over all isotopes with $2Z \neq A$.
In order to get this value and the density profiles,
the precollapse models of massive stars of Woosley \& Weaver (WW)
\cite{WW1995} and Nomoto \& Hashimoto (NH) \cite{NH1988} are used.
We assume that the dynamical effect can be ignored within the 
time scale of the neutrino emission, and hence use the above
static models. The mass and radius of the
helium core of a $15 M_\odot$ main sequence star is $\sim
4 M_\odot$ and $\sim 1 R_\odot$, respectively, and
its free-fall time scale, $(\sqrt{G \rho})^{-1}$ is $\sim
10^2$--$10^3$ [sec], which is longer than the neutrino emission
time scale (at most a few tens of seconds). It takes about 
several tens of seconds for the shock wave generated at the core
bounce to reach the hydrogen envelope \cite{WW1995}, and 
the inner region of the isotopically neutral region may be
disturbed by the shock wave. We will discuss about this in section
\ref{sec:discussion}.
The calculation of the WW models of 15 and 25 $M_\odot$ (hereafter
WW15 and WW25, respectively) includes 200 isotopes, up to $^{71}$Ge. 
Although the network of 19 isotopes
is used for energy generation up to the end of oxygen burning,
the network of 200 isotopes is updated in each cycle and mixed using
the same diffusion coefficients.
The NH models are 4 and 8 $M_\odot$ helium cores (hereafter NH4 and NH8,
respectively) corresponding approximately to 15 and 25 $M_\odot$
main sequence stars. Their calculation includes 30 isotopes 
up to the end of oxygen burning, which are also used for the energy
generation. The WW and NH models use the different reaction rates of 
$^{12}{\rm C} (\alpha , \gamma) ^{16}{\rm O}$, and the treatment of convection
is also different. As for the WW models, we use 
the models with two different metallicities: the solar and zero
metallicity, and the metal abundance of the NH models is that of the
Sun. 

By using the data of composition as well as the density profile
of the solar metallicity
WW models (WW15S and WW25S, where `S' denotes the solar metallicity), 
$|\Delta V|$ in the WW15S and WW25S models are
depicted in Figs. \ref{fig:WW15S-ham}
and \ref{fig:WW25S-ham}, respectively,
by the thick solid lines as a function of the radius from the center
of the star.
Also shown by the dashed lines is $|\Delta H|$ when 
$\Delta m^2 / E_\nu$ is $10^{-4}$ and $10^{-6}$ [eV$^2$/MeV].
The dominantly existent nuclei are
also indicated for each layer in the top of these figures.
In the neutronized iron core,
$Y_e$ is smaller than 0.5, and $\Delta V$ is positive
and much larger than $\Delta m^2 / E_\nu$ unless $\Delta m^2 / E_\nu$
is larger than $10^{-1}$ [eV$^2$/MeV].
We consider the range of $\Delta m^2 / E_\nu$ below this 
value, and hence the resonance does not occur in the iron core.
Above the iron core, i.e., in the isotopically
neutral region, $Y_e$ becomes quite close to 0.5 (still $Y_e < 0.5$)
and $\Delta V$
is strongly suppressed, typically by a factor of $\sim 10^{-3}$
in solar metallicity stars,
and the term $\mu_\nu B$ becomes
more effective. This suppression continues to the end of the 
isotopically neutral region, namely, just below the hydrogen
envelope. 
The isotopically neutral region is roughly divided into 
the four layers: O+Si, O+Ne+Mg, O+C, and He layer, 
from inner to outer region. 
The values of $(1-2Y_e)$ and some nuclei which are relevant to
the deviation of $Y_e$ from 0.5 are tabulated in Table 
\ref{table:ye} for each layer and for the six precollapse models
used in this paper.
For the solar metallicity models, $(1-2Y_e)$ is determined mainly by
the isotopes such as $^{22}$Ne, $^{25,26}$Mg, $^{27}$Al, $^{34}$S,
$^{38}$Ar, and so on.

The resonance occurs when $\Delta V$ becomes smaller than $\Delta m^2 /
2 E_\nu$, and after the resonance (above the resonance layer)
$\Delta H$ becomes constant with radius
because $\Delta V$ is negligibly small. If the strength of the magnetic
field is sufficiently strong for the satisfaction of the adiabaticity
condition at the resonance layer, 
the neutrinos are resonantly converted into the other helicity state.
The magnetic fields and the adiabaticity condition are discussed in 
the following subsection.
The resonance layer is in the isotopically neutral region
if $\Delta m^2 / E_\nu$ is in the range of roughly
$10^{-10}$--$10^{-1}$ [eV$^2$/MeV] (slightly dependent on the 
stellar models), and the resonance layer moves inward with increasing
$\Delta m^2 / E_\nu$.
If $\Delta m^2 / E_\nu$ is smaller than $10^{-10}$ [eV$^2$/MeV], 
the mass term has no effect on $\Delta H$ in this region of the
solar metallicity models, 
and the resonance occurs at the boundary between the helium
layer and the hydrogen envelope due to the change of the sign of
($1-2Y_e$). In contrast with the flavor oscillation,
the matter potential changes its sign by itself and the resonance can
occur without the mass term, $\Delta m^2 / E_\nu$.
However, as explained in the next subsection,
if a dipole moment is assumed as the
global structure of the magnetic fields, it seems difficult
that $B$ is strong enough to satisfy the adiabaticity condition at
this boundary in the solar metallicity models. 
In the hydrogen envelope, $Y_e$ is about 0.8 and 
the suppression of $(1-2Y_e)$ does not work any more. We can see that
most of the qualitative features are the same for the two models:
WW15S and WW25S, and the dependence on the stellar masses is rather
small. Note that our result gives 1--2 orders of magnitude larger 
$\Delta V$ than that in 
the earlier calculation by Athar et al.\cite{Athar-etal}, in which
the older 15 $M_\odot$ Woosley \& Weaver model \cite{WW1986,%
Woosley-Langer-Weaver} is used.
It is probably because our calculation of $(1-2Y_e)$ includes
the larger network of isotopes used in the latest WW models.

In Figs. \ref{fig:NH4-ham} and \ref{fig:NH8-ham}, we show the
same with Figs. \ref{fig:WW15S-ham} and \ref{fig:WW25S-ham},
but for the Nomoto \& Hashimoto models. It can be seen that
the profiles of $\Delta V$ of the 
WW and NH models are not so different, and the model dependence seems 
rather small. However, the situation is drastically changed when 
we consider the effect of different metallicities.
Figs. \ref{fig:WW15Z-ham} and \ref{fig:WW25Z-ham} are the same with
Figs. \ref{fig:WW15S-ham} and \ref{fig:WW25S-ham}, but for the 
zero metallicity WW models: WW15Z and WW25Z. (`Z' denotes the 
zero metallicity.) In the O+Si and O+Ne+Mg
layers, $(1-2Y_e)$  is smaller than that of the solar metallicity
models by about 1 order of magnitude,
and in the O+C and He layers, $(1-2Y_e)$ is
further strongly suppressed (4--6 orders of magnitudes)
because of the lack of the heavy nuclei which cause the deviation
of $Y_e$ from 0.5 (Table \ref{table:ye}).
In consequence, the metallicity effect becomes
especially important when $\Delta m^2 / E_\nu$ is smaller than $\sim
10^{-6}$ [eV$^2$/MeV]. 
How this effect changes the profile of
the conversion probability will be discussed in more detail in 
section \ref{sec:results}.

\subsection{Magnetic Fields}
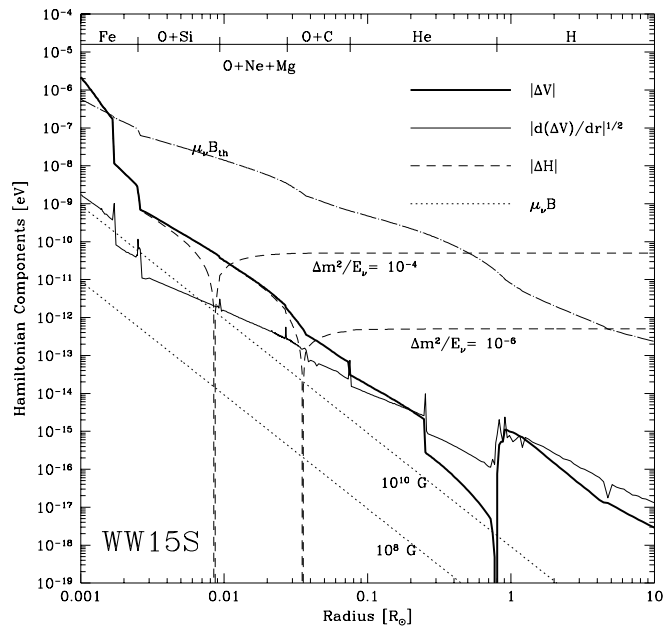
\begin{figure}
\caption{The components of the Hamiltonian in the propagation equation 
(\protect\ref{eq:schrodinger}) are shown for the Woosley \&
Weaver's solar-metallicity 15 $M_\odot$ model, 
as a function of the radius from the center of the star.
The absolute value of the effective matter potential, $\Delta V$,
is shown by the thick, solid line. The absolute value of 
the difference of the diagonal components
of the Hamiltonian, $\Delta H$, is shown by the dashed lines for 
the two values of the mass term, $\Delta m^2 / E_\nu = 10^{-4}$
and $10^{-6}$ [eV$^2$/MeV]. The resonant conversion occurs where
$\Delta H = 0$ (the resonance layer).
The dotted lines show $\mu_\nu B$,
i.e., the off-diagonal components of the Hamiltonian.
The indicated values of $B$ are those
at the surface of the iron core ($B_0$) in units
of Gauss, assuming $\mu_\nu = 10^{-12}\mu_B$. The thin solid line
shows $|d(\Delta V)/dr|^{1/2}$ [eV], and the adiabaticity condition is 
satisfied when $\mu_\nu B \protect\agt |d(\Delta V)/dr|^{1/2}$
at the resonance layer. The dot-dashed
line shows $\mu_\nu B_{th}$, again assuming $\mu_\nu = 10^{-12}\mu_B$,
where $B_{th}$ is the magnetic field
whose energy density is the same with the thermal energy density in
the star. The dominantly existent nuclei in each layer of 
the precollapse model are shown in the top of this figure.}
\label{fig:WW15S-ham}
\end{figure}

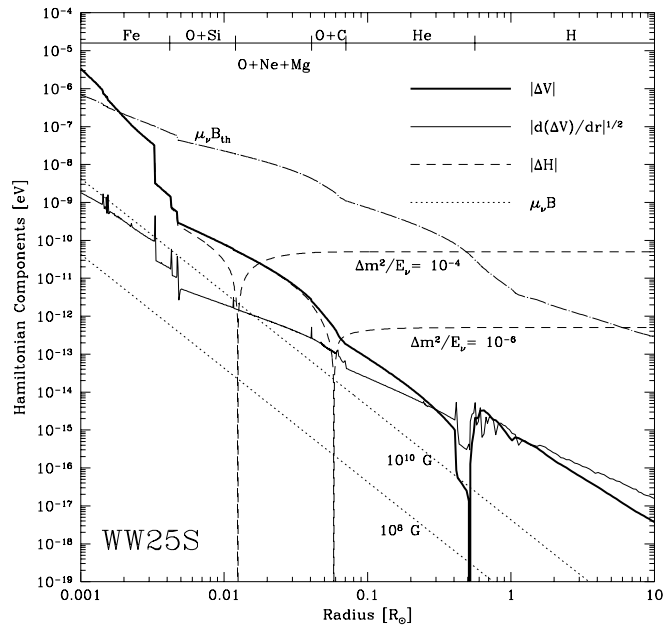
\begin{figure}
\caption{The same as  Fig. \protect\ref{fig:WW15S-ham}, but for the 
Woosley \& Weaver's solar-metallicity 25 $M_\odot$ model.}
\label{fig:WW25S-ham}
\end{figure}

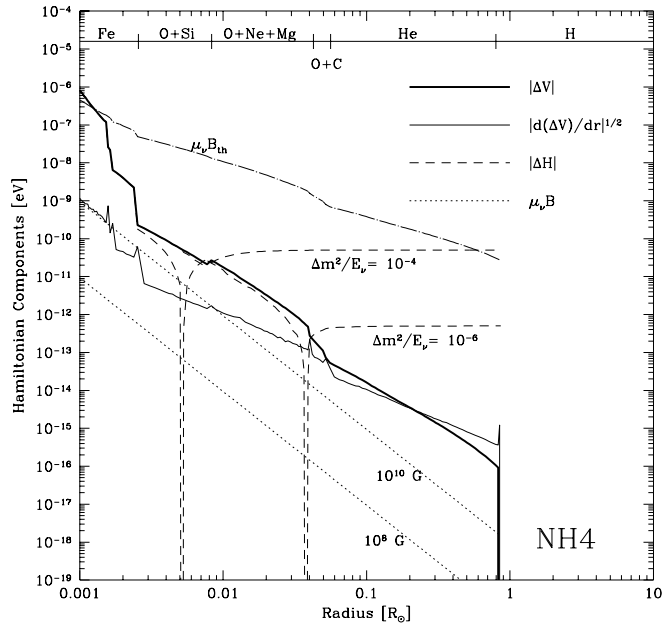
\begin{figure}
\caption{The same as Fig. \protect\ref{fig:WW15S-ham}, but for the 
Nomoto \& Hashimoto's 4 $M_\odot$ helium-core model.
}
\label{fig:NH4-ham}
\end{figure}

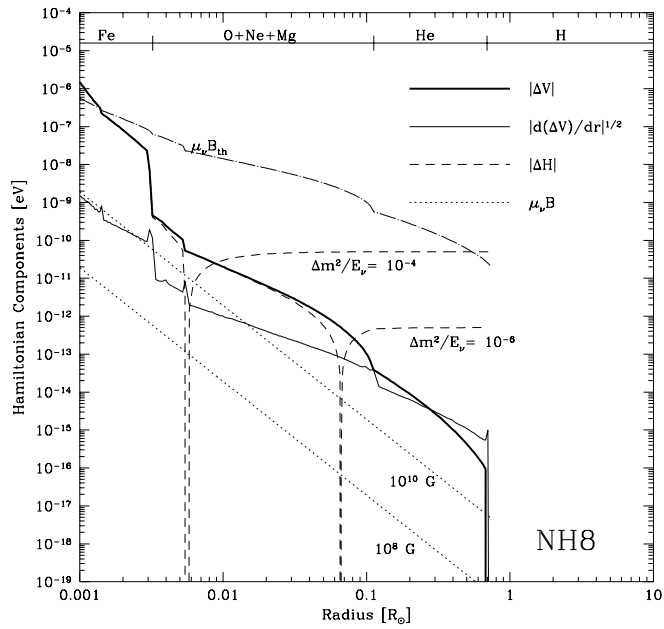
\begin{figure}
\caption{
The same as Fig. \protect\ref{fig:WW15S-ham}, but for the Nomoto \&
Hashimoto's 8 $M_\odot$ helium-core model.
The O+Si and O+C layers are negligibly thin in this figure.
}
\label{fig:NH8-ham}
\end{figure}

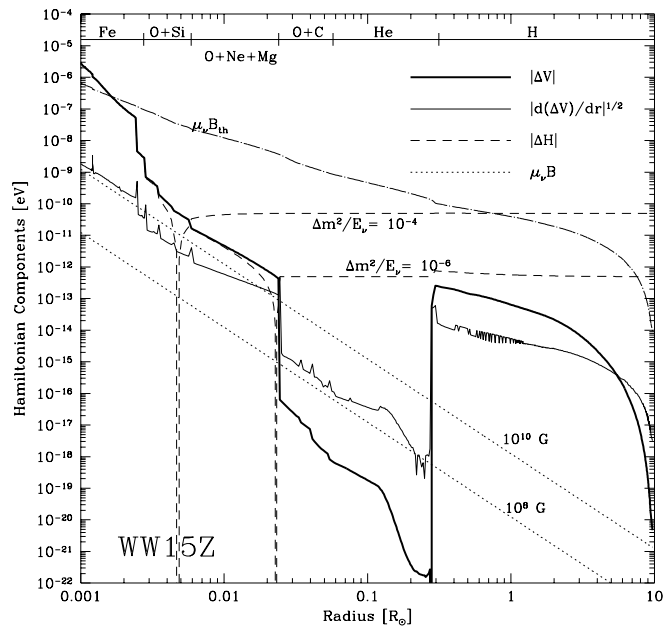
\begin{figure}
\caption{The same as Fig. \protect\ref{fig:WW15S-ham}, 
but for the Woosley \& Weaver's zero-metallicity 15 $M_\odot$ model. 
Note that the scale of the vertical axis
is different from the previous four figures.}
\label{fig:WW15Z-ham}
\end{figure}

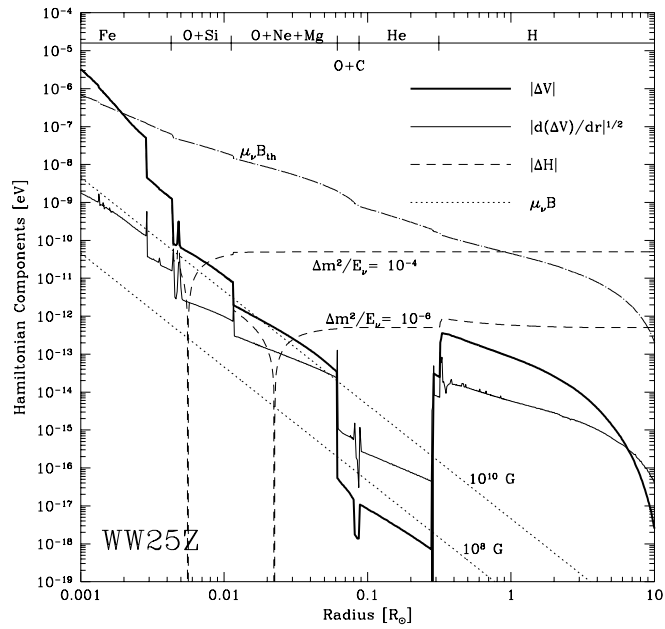
\begin{figure}
\caption{The same as Fig. \protect\ref{fig:WW15S-ham}, 
but for the Woosley \& Weaver's zero-metallicity 25 $M_\odot$ model.}
\label{fig:WW25Z-ham}
\end{figure}

Let us consider the magnetic fields in the isotopically
neutral region. 
In the earlier publication\cite{Athar-etal}, the strength of
magnetic fields was normalized at the surface of the newly born
neutron star ($r \sim$ 10 km), but it is unlikely that the 
magnetic fields of a nascent neutron star have some effects on
the far outer region, such as the isotopically neutral region,
within the short time scale of the neutrino burst.
The magnetic fields should be normalized by the fields which 
are static and 
existent before the core collapse. 
The strength of such magnetic fields above the 
surface of the iron core may be inferred from that observed on the 
surface of white dwarfs, because the iron core of giant stars
is similar to white dwarfs in the point that both are 
sustained against the gravitational collapse by the degenerate
pressure of electrons. The observations of the magnetic fields
in white dwarfs show that the strength 
spreads in a wide range of $10^7$--$10^9$ Gauss\cite{Chanmugam}.
Taking account of the possibility of the decay of magnetic fields
in white dwarfs, it is not unnatural to consider the magnetic fields
up to $10^{10}$ Gauss at the surface of the iron core.
As for the global structure of the fields, although the optimistic estimate
of $B \propto r^{-2}$ is sometimes discussed
from the argument of the flux freezing,
a magnetic dipole is natural as static and global fields;
we hence assume such fields in this paper. Therefore the off-diagonal
element of the Hamiltonian in Eq.(\ref{eq:schrodinger}), 
$\mu_\nu B_\bot$ becomes $\mu_\nu B_0 (r_0/r)^3 \sin \Theta$,
where $B_0$ is the strength of the magnetic field at the equator
on the iron core surface, $r_0$ the radius of the iron core,
and $\Theta$ the angle between the pole of the magnetic dipole and 
the direction of neutrino propagation.
If the magnetic field is normalized at the surface of the neutron
star, the radial dependence of a dipole ($\propto r^{-3}$) gives
too small field in the isotopically neutral region, but
the normalization at the surface of the 
iron core inferred from the observations
of white dwarfs makes it possible that the magnetic field is
sufficiently strong in the isotopically neutral region
under the condition of a global dipole moment.
The lines of $\mu_\nu B$ [eV] are shown in Figs.
\ref{fig:WW15S-ham}--\ref{fig:WW25Z-ham} for $B_0 = 
10^8$ and $10^{10}$ [Gauss], assuming $\mu_\nu = 10^{-12} \mu_B$.
The strength of magnetic fields is also discussed from the
argument of energetics. The energy density of
the maximum strength of magnetic
fields should at most be the same order of magnitudes
with that of the thermal plasma in the star.
Let us define the magnetic fields $B_{th}$,
whose energy density is the same
with that of the gas in the star:
\begin{equation}
\frac{1}{8\pi} B_{th}^2 = \frac{3}{2}\frac{\rho}
{\tilde{\mu} m_p}kT \ ,
\end{equation}
where $\rho$ is the density, $T$ temperature, $k$ the Bolzman
constant, and $\tilde{\mu}$ the mean molecular weight.
The line of $\mu_\nu B_{th}$ is depicted in Figures 
\ref{fig:WW15S-ham}--\ref{fig:WW25Z-ham}, assuming $\tilde{\mu} = 1$
and $\mu_\nu = 10^{-12} \mu_B$,
and it can be seen that 
the magnetic fields up to $B_0 \sim 10^{10}$ [Gauss] are
far below $B_{th}$ and therefore 
natural from the view point of energetics.

If there is no matter potential, the complete precession of 
$\nu_R \leftrightarrow \nu_L$ occurs; however, the precession
amplitude is suppressed by the matter potential.
The precession amplitude is given in the form\cite{Athar-etal,VVO}:
\begin{equation}
A_p = \frac{(2\mu_\nu B)^2}{(2\mu_\nu B)^2 + (\Delta H)^2} \ .
\label{eq:precession-amplitude}
\end{equation}
In the neutronized iron core, $\Delta H$ is much larger than
$\mu_\nu B$ even when $B_0 \sim 10^{10}$ [Gauss], and the precession
below the surface of the iron core can be completely neglected.
(In other words, we can start the calculation with the pure neutrino
states from the iron core surface, with $B_0 \alt 10^{10}$ [Gauss].)
Above the iron core, i.e., in the isotopically neutral region
in the solar metallicity models,
$\mu_\nu B$ is still much lower than $\Delta V$ (or $\Delta H$), 
except at the resonance
layer or the boundary of the helium layer and the hydrogen envelope,
as shown in Figs. \ref{fig:WW15S-ham}--\ref{fig:NH8-ham}.
Therefore, the precession does not occur in 
the solar metallicity stars.
However, if the strength of the magnetic fields is strong enough
to satisfy the adiabaticity condition at the resonance layer,
neutrinos are resonantly converted into other types of neutrinos. 
The adiabaticity condition is satisfied 
when the precession length at the resonance layer, $(\mu_\nu B)^{-1}$,
is shorter than the thickness of the resonance layer, i.e.,
\begin{equation}
\mu_\nu B \agt \left|\frac{d(\Delta H)}{dr}
\right|^{\frac{1}{2}} =
\left|\frac{d(\Delta V)}{dr}
\right|^{\frac{1}{2}} \  ({\rm at \ the \ resonance}),
\label{eq:adiabaticity-cond}
\end{equation}
since the thickness of the resonance layer, $\Delta r_{\rm res}$,
is given as 
\begin{equation}
\Delta r_{\rm res} = \mu_\nu B \left( \left| \frac{d(\Delta H)}{dr}
\right| \right)^{-1} \ . 
\end{equation}
Note that 
the suppression of $(1-2Y_e)$ in the isotopically neutral region
makes the adiabaticity condition well satisfied because it reduces
the right hand side of Eq.(\ref{eq:adiabaticity-cond}) by a factor
of $(1-2Y_e)^{1/2}$.
In order to show how this condition is satisfied,
$|d(\Delta V)/dr|^{1/2}$ is shown by the thin solid lines 
in Figs. \ref{fig:WW15S-ham}--\ref{fig:WW25Z-ham}. 
If $\mu_\nu B$ is (roughly) larger than 
$|d(\Delta V)/dr|^{1/2}$ at the resonance layer
($\Delta H = 0$), the adiabaticity
condition is satisfied and $\bar\nu_e$'s and $\nu_\mu$'s are 
mutually converted.
In both the WW and NH models with the solar metallicity,
the region where this condition is 
satisfied appears with $B_0 \agt 10^{10}$ [Gauss].
Because the slope of 
$|d(\Delta V)/dr|^{1/2}$ is flatter than that of 
$\mu_\nu B$, this condition is 
satisfied better in the inner region of the star, in other words,
with large values of $\Delta m^2 / E_\nu$, in the solar metallicity
models. When $\Delta m^2 / E_\nu $ is smaller than
$\sim 10^{-10}$ [eV$^2$/MeV] and the resonance layer
lies at the boundary of the helium layer and the hydrogen envelope,
unnaturally strong 
magnetic fields are necessary for satisfaction of the adiabaticity
condition. However, in the zero metallicity stars, because the value of 
$(1-2Y_e)$ is very strongly suppressed in the O+C and He layers,
$\mu_\nu B$ becomes much larger than $\Delta H$ and hence
the strong precession between different chiralities occurs with
small $\Delta m^2 / E_\nu$
(Figs. \ref{fig:WW15Z-ham} and \ref{fig:WW25Z-ham}).
Since the adiabaticity may be broken at the quite large jump of
the matter potential at the boundary of the helium layer and the hydrogen
envelope, the detailed calculation is necessary for the conversion probability
when $\Delta m^2 / E_\nu \alt 10^{-8}$ [eV$^2$/MeV].
It is apparent that the conversion probability in zero metallicity stars
will be completely different from the solar metallicity stars.
Now all of the qualitative features of the conversion can be understood from 
Figs. \ref{fig:WW15S-ham}--\ref{fig:WW25Z-ham}
and the results of final conversion probability obtained by
solving the evolution equation (Eq. \ref{eq:schrodinger}) numerically
are given in the following section.

\section{Results}
\label{sec:results}
\subsection{Conversion Probability Maps}
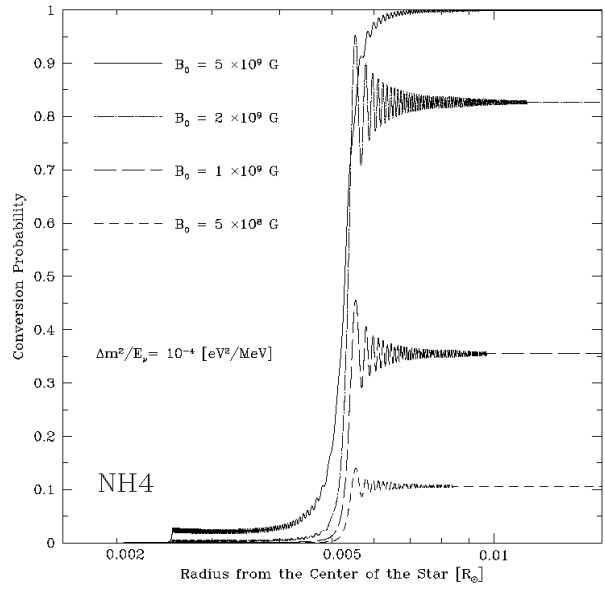
\begin{figure}
\caption{
The evolution of conversion probability with the radius from
the center of the star. The Nomoto \& Hashimoto's 4 $M_\odot$
helium-core model is used,
and $\Delta m^2 / E_\nu$ is assumed to be $10^{-4}$[eV$^2$/MeV].
The four different values of $B_0$ are used for the calculation,
where $B_0$ is the magnetic field 
at the surface of the iron core. The neutrino magnetic moment,
$\mu_\nu$, is assumed to be $10^{-12} \mu_B$.       
}
\label{fig:demo-1}
\end{figure}

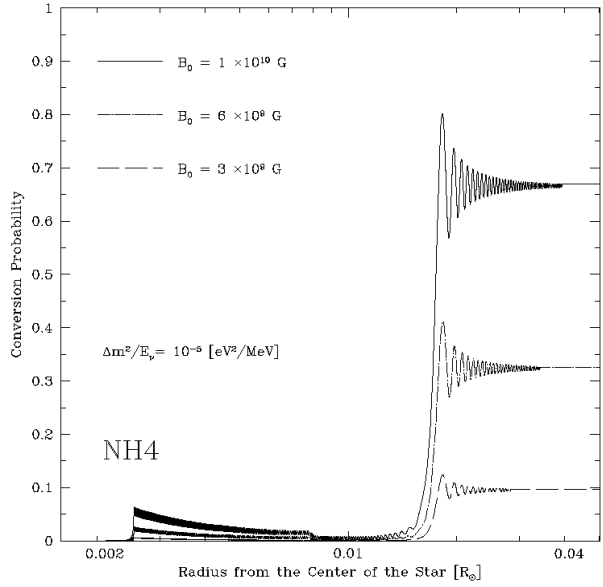
\begin{figure}
\caption{
The same as Fig. \protect\ref{fig:demo-1}, 
but for the case with $\Delta m^2 / E_\nu = 10^{-5}$[eV$^2$/MeV].
The three values of $B_0$ are used as indicated in the figure.
}
\label{fig:demo-2}
\end{figure}

In this section the contour maps of the conversion
probability ($\bar\nu_e \leftrightarrow \nu_\mu$)
are given for all the models used in this paper
as a function of $\Delta m^2 / E_\nu$ and $B$
at the surface of the neutronized iron core ($B_0$).
Before we proceed to contour maps, 
the evolution of conversion probability in the 
isotopically neutral region along the trajectory of
neutrinos is shown for some cases as a demonstration.
Fig. \ref{fig:demo-1} shows the conversion probability as a 
function of radius from the center of the star 
using the model NH4  for some
values of $B_0$, with
$\Delta m^2 / E_\nu = 10^{-4}$[eV$^2$/MeV],
$\mu_\nu = 10^{-12}\mu_B$, and $\cos 2 \theta = 1$.
The resonance layer lies at $r \sim 5 \times 10^{-3} R_\odot$
in the O+Si layer and its location
is never changed by strength of the magnetic fields.
We can see in this figure that the conversion probability
becomes larger with increasing strength of magnetic fields, and 
the complete conversion occurs with the magnetic fields strong
enough ($B_0 \agt 5 \times 10^9$ [Gauss], in this case)
to satisfy the adiabaticity condition (see also Figure
\ref{fig:NH4-ham}). Fig. \ref{fig:demo-2} is the same with Fig.
\ref{fig:demo-1},
but $\Delta m^2 / E_\nu$ is $10^{-5}$[eV$^2$/MeV] and 
the resonance layer is hence in more outer region at
$r \sim 1.5 \times 10^{-2} R_\odot$ (O+Ne+Mg layer).
As shown in these figures, the necessary $B_0$ for the complete
conversion becomes larger with decreasing $\Delta m^2 / E_\nu$
in the solar metallicity models, because the adiabaticity condition
is well satisfied with larger $\Delta m^2 / E_\nu$, 
as discussed in the previous section.
In both the figures, the conversion probability
jumps up a little at the radius of about 2.5 $\times 10^{-3} R_\odot$,
because this radius corresponds to the surface of the iron core and
the value of ($1-2Y_e$) drops quite suddenly here.

Now we calculate the contour maps of the conversion probability
for the solar metallicity models, in the region of 
$\Delta m^2 / E_\nu = 10^{-8}$--$10^{-1}$[eV$^2$/MeV] and 
$B_0 = 10^8$--$10^{10}$ [Gauss],
and the results are given in Figs.
\ref{fig:WW15S-contour}--\ref{fig:NH8-contour} (for WW15S, WW25S,
NH4, and NH8, respectively). In the region of 
$\Delta m^2 / E_\nu < 10^{-8}$[eV$^2$/MeV] or $B_0 < 10^8$ [Gauss],
the conversion does not occur because of too weak magnetic fields.
Magnetic fields stronger than $10^{10}$ [Gauss] induce 
the precession below the surface of 
the iron core which cannot be ignored,
and $\Delta m^2 / E_\nu$ larger than
$10^{-1}$[eV$^2$/MeV] leads to the resonance below the
surface of the iron core. In this paper, we consider the parameter
region in which the conversion or precession 
below the iron core surface can be neglected.
The contours are depicted with the probability intervals of 0.1, assuming
$\mu_\nu = 10^{-12} \mu_B$ and $\cos 2 \theta = 1$.
It can be seen that some observable effects on the 
spectrum of the emitted $\bar\nu_e$'s are expected if $B_0$ is
stronger than $\sim$10$^9$ [Gauss] and $\Delta m^2$ is
larger than $\sim$10$^{-5}$ [eV$^2$]. Note that the typical energy
range of the neutrinos which are observed in a water \v{C}erenkov detector
is 10--70 MeV. The lower margins of the strong conversion region 
($P >$ 0.9, where $P$ is the conversion probability) in
the contour maps are, in all the four models,
contours which runs from the upper left to the lower right
direction. This is due to the fact that the adiabaticity condition is 
well satisfied with larger values of $\Delta m^2 / E_\nu$,
as discussed in the previous section.
We refer to this marginal region in the contour maps as ``the 
continuous deformation region'', because the conversion probability
continuously decreases with increasing neutrino energy in this region
and the spectral deformation is expected to be continuous.
What is interesting about these maps
is that some band-like patterns can be 
seen in the relation of the conversion 
probability and the value of $\Delta m^2 /
E_\nu$. For example, the conversion probability in the region of 
$\Delta m^2 / E_\nu = 5 \times 10^{-4}
$--$5 \times 10^{-3}$ [eV$^2$/MeV] in Fig. \ref{fig:NH4-contour} 
is much lower than that in the other regions of the map. These patterns
come directly from the jumps in the matter potential due to
the onion-like structure of the isotopic composition
in giant stars\cite{Athar-etal}. 
When $\Delta m^2 / E_\nu$ is in the above region,
the resonance in the model NH4
occurs at the surface of the iron core
and the interval of $\Delta m^2 / E_\nu$
corresponds to the jump of $\Delta V$ 
at the surface (see Fig. \ref{fig:NH4-ham}). Since the matter potential
changes suddenly here, very strong magnetic field is necessary
for the satisfaction of the adiabaticity condition, and consequently
the resonant conversion is significantly suppressed.
We refer to such bands
of $\Delta m^2 / E_\nu$ as ``the weak adiabaticity band'', 
hereafter. At each boundaries of the onion-like structure of 
massive stars,
this weak adiabaticity band appears due to the jump in the 
matter potential.
It should also be noted that in the weak adiabaticity bands,
the conversion probability only weakly depends on $\Delta m^2 /
E_\nu$ because the location of the resonance layer is not changed
in a band. In Figs. \ref{fig:WW15S-contour}--\ref{fig:NH8-contour},
we can see the difference between the WW and NH models
as well as between 15 and 25 $M_\odot$ models.
Although there are some quantitative differences,
almost all qualitative features are the same
for these four models. 

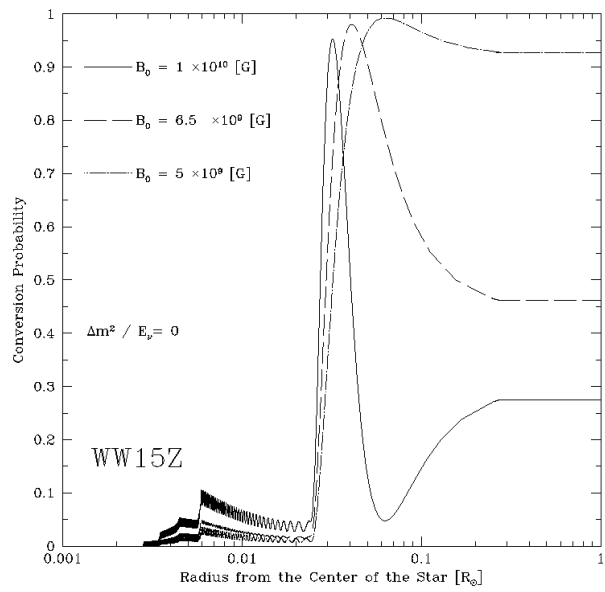
\begin{figure}
\caption{ The same as Fig. \protect\ref{fig:demo-1}, but for the Woosley \&
Weaver's zero-metallicity 15 $M_\odot$ model and $\Delta m^2 / E_\nu = 0$.
The three values of $B_0$ are used as indicated in the figure.
Strong precession occurs in the region of 0.025--0.28 $R_\odot$.}
\label{fig:demo-3}
\end{figure}

Next we show the contour maps of the conversion probability
for the zero metallicity models, in Figs. \ref{fig:WW15Z-contour}
and \ref{fig:WW25Z-contour} (for the models WW15Z and WW25Z, 
respectively), with the region of $\Delta m^2 / E_\nu = 10^{-11}$%
--$10^{-1}$ [eV$^2$/MeV] and $B_0 = 10^8$--$10^{10}$.
When $\Delta m^2 / E_\nu$ is larger than $\sim 10^{-6}$
[eV$^2$/MeV], the profile of the contour maps is qualitatively similar
to that of the solar metallicity models. But the
adiabaticity condition can be satisfied with smaller strength
of magnetic fields and the region of complete conversion becomes
somewhat larger, because in the inner part of the isotopically neutral
region (O+Si and O+Ne+Mg layers) the value of $(1-2Y_e)$
is about 1 order of magnitude smaller than that in the solar 
metallicity models. Especially, in the model WW25Z, $\Delta H$ and 
$\mu_\nu B$ are comparable in this region (see Fig. \ref{fig:%
WW25Z-ham}) and the precession effect is no longer negligible,
leading to the more complicated feature of the contour map of WW25Z
than of WW15Z.
When $\Delta m^2 / E_\nu \alt
10^{-6}$ [eV$^2$/MeV], the strong precession occurs in the
outer part of the isotopically neutral region (O+C and He layers),
where $\mu_\nu B$ is much higher than $\Delta H$.
In contrast to 
the solar metallicity models, the conversion still occurs with such a low 
value of $\Delta m^2 / E_\nu$, even when $\Delta m^2 = 0$.
Further interesting is that with $\Delta m^2 / E_\nu$ lower than
$\sim$10$^{-6}$ [eV$^2$/MeV], the conversion probability
changes periodically with $\mu_\nu B_0$ (Fig. \ref{fig:%
WW15Z-contour}). This can be understood as follows. 
The precession effect in the outer part of the isotopically neutral
region is very profound and then this precession is stopped almost suddenly
at the boundary of the helium layer and the hydrogen envelope 
where $|\Delta V|$
increases by 5--10 orders of magnitude. The final phase of the precession
strongly depends on $\mu_\nu B_0$,
because the precession length is given as
\begin{equation}
L = \frac{\pi}{\sqrt{\left( \frac{\Delta m^2}{4 E_\nu}
\right)^2 + \left( \mu_\nu B \right)^2}} \ .
\label{eq:osc-length}
\end{equation}
Therefore the conversion probability
changes periodically with $\mu_\nu B_0$.
The examples of strong precession effect are shown in Fig.
\ref{fig:demo-3}, using the WW15Z model with $\Delta m^2 = 0$ 
and some values of $B_0$.
The conversion probability as a function of the radius is shown 
in this figure.
The precession begins at $r = 0.025 R_\odot$ and ceases at $r =
0.28 R_\odot$ (see also Fig. \ref{fig:WW15Z-ham}). One can see 
that the precession length becomes larger with propagation of
neutrinos, because $B$ decreases with $r$. It is also clear that 
the change in $B_0$ leads to the change of the precession length,
and hence to the oscillation of the final phase of precession.
Below $\Delta m^2 / E_\nu \sim 10^{-11}$
[eV$^2$/MeV], the effect of the mass term can be completely neglected
and the conversion probability becomes constant (but never vanishes)
with neutrino mass or energy.

\subsection{Spectral Deformation}
All of the qualitative features of the spectral deformation
due to the resonant spin-flavor conversion of $\bar\nu_e \leftrightarrow
\nu_\mu$ are clearly understood by 
the contour maps given in the previous section (Figs.
\ref{fig:WW15S-contour}--\ref{fig:WW25Z-contour}). 
The most easily detectable flavor in a water \v{C}erenkov detector
is $\bar\nu_e$'s because of the large cross section of
the reaction $\bar\nu_e p \rightarrow n e^+$, and 
they are detectable 
above the positron energy of $\sim$5 MeV in the Super-Kamiokande
detector, which has the fiducial volume of 22,000 tons \cite{SK}. 
If we consider the positron energy range of 
10--70 MeV, which includes almost all of the events,
this range corresponds to a vertical bar in the contour maps
with fixed values of $\Delta m^2$ and $B_0$. The samples of such
a bar are shown in Figs. 
\ref{fig:NH4-contour} and \ref{fig:NH8-contour} (NH models), and
the corresponding spectral deformation of the events at the SK
are shown in Fig. \ref{fig:spec-def-1}.
The distance of the supernova is set to 10 kpc and 
the total energy of each type of neutrinos is assumed to be 
$5 \times 10^{52}$ erg. We use 5 and 8 MeV as 
the temperature of $\bar\nu_e$'s and $\nu_\mu$'s,
respectively, and the Fermi-Dirac distribution with zero chemical
potential is assumed for both $\bar\nu_e$'s and 
$\nu_\mu$'s\cite{SNneu-property}.
The cross section of the dominant reaction of $\bar\nu_e
p \rightarrow n e^+$ is $9.72 \times 10^{-44} 
E_e p_e$ cm$^2$\cite{SK},
where $E_e$ and $p_e$ is the energy 
and momentum of recoil positrons. The appropriate detection efficiency
curve is also taken into account\cite{efficiency}. The thick
solid line in Fig. \ref{fig:spec-def-1}
is the expected differential event number
of $\bar\nu_e$'s without any oscillation or conversion.

As mentioned in the previous section, the three characteristic
regions appear in the contour maps for the solar metallicity
models: A) the complete conversion
region, B) the continuous deformation region, and C) the weak adiabaticity
band. (A, B, and C correspond to those in Figs.
\ref{fig:NH4-contour} and \ref{fig:NH8-contour}.)
When the conversion is complete, we can see the original
$\nu_\mu$ spectrum as $\bar\nu_e$'s, and the event number 
is considerably enhanced because of the higher average energy
(thin solid line in Fig. \ref{fig:spec-def-1}).
When the vertical line in the contour map lies in the continuous
deformation region, conversion probability decreases with increasing
energy of neutrinos and consequently the original $\nu_\mu$'s
are dominant in the lower energy range, while the original $\bar\nu_e$'s
are dominant in the higher energy range (short-dashed line in Fig.
\ref{fig:spec-def-1}, 
also corresponding to the vertical line (B) in Fig. 
\ref{fig:NH8-contour}). 
Note that this feature is based upon the assumption
that the radial dependence of the magnetic fields is a dipole ($B
\propto r^{-3}$) and $\mu_\nu B$ drops faster than 
$|d(\Delta V)/dr|^{1/2}$
with increasing radius. On the other hand, in the weak adiabaticity 
band, the energy dependence of conversion probability is rather weak
(long-dashed line in Fig. \ref{fig:spec-def-1}, 
also corresponding to
the vertical line (C) in Fig. \ref{fig:NH4-contour}). 
Because the resonance always
occurs in the same place (jumps in the matter potential), 
this feature does not depend on the assumption of the radial
dependence of magnetic fields, in contrast to the case (B).
Finally, quite interesting deformation is expected if the vertical
line in the contour maps crosses the boundary of the weak adiabaticity
band (the vertical line (D) in Fig. \ref{fig:NH4-contour}). Because 
the conversion probability changes almost suddenly at the boundary,
the spectrum of the event rate suffers drastical
deformation at a certain positron energy (dot-dashed line in Fig.
\ref{fig:spec-def-1}). 
The used values of ($\Delta m^2$
[eV$^2$], $B_0$ [Gauss]) for 
the vertical lines A, B, C, and D in Figs. \ref{fig:NH4-contour}
and \ref{fig:NH8-contour} are (5$\times 10^{-3},
7\times 10^9$), ($5\times 10^{-4}, 5\times 10^9$), 
($3\times 10^{-2}, 5\times 10^9$), and ($1.5\times 10^{-2}, 2\times
10^9$), respectively.

In the zero metallicity models, the feature of the spectrum
deformation is similar to that of the solar metallictiy models
when $\Delta m^2 / E_\nu \agt 10^{-7}$ [eV$^2$/MeV].
However, if $\Delta m^2 \sim 10^{-7}$--$10^{-6}$ [eV$^2$], 
the conversion probability increases with neutrino energy, because
the precession in the outer part of the isotopically neutral region
becomes effective (Fig. \ref{fig:WW15Z-contour}). With $\Delta m^2 / 
E_\nu \alt 10^{-9}$ [eV$^2$/MeV], the conversion probability
becomes constant with neutrino energy, but changes
periodically with $B_0$. In the model WW25Z, the precession
in the inner part of the isotopically neutral region (O+Si and 
O+Ne+Mg layers) is also effective, and the probability may change rapidly
and complicatedly with neutrino energy (Fig \ref{fig:WW25Z-contour}).

\section{Discussion}
\label{sec:discussion}
We found that the difference of the stellar metallicity
significantly affects the resonant spin-flavor
conversion of $\bar\nu_e \leftrightarrow \nu_\mu$,
and some implications from this fact are discussed in the following.
The lifetime of massive stars which end their life by the 
gravitational collapses is very shorter than that of the Sun,
and the progenitors of observed supernovae are therefore younger.
Consequently, the metallicity of the Galactic
supernova is expected to be at least the solar abundance or more metal-rich.
If the metallicity is higher than that of the Sun,
the suppression of $(1-2Y_e)$ will be weaker and the $B_0$ which is required
to satisfy the adiabaticity condition becomes larger.
On the other hand, the Large and Small Magellanic Clouds are
known to be very metal-poor systems\cite{MC-metallicity}.
Therefore, the resonant conversion will occur with smaller
magnetic fields in supernovae in the Magellanic Clouds, and
also the precession effect may be observed. The another object which
has relation to the metallicity effect is the supernova relic neutrino
background (SRN)\cite{SRN}. Because the SRN is the accumulation
of neutrinos from supernovae which have ever occurred in the universe,
the SRN includes neutrinos from supernovae with quite low metallicity
in the early phase of galaxy formation. The conversion of
$\bar\nu_e \leftrightarrow \nu_\mu$  conpensates the energy
degradation due to the cosmological redshift effect and enhances
the expected event rate of the SRN. 
However, because of the small expected event rate at the SK%
\cite{SRN}, it will be difficult to get some decisive information 
on the spin-flavor conversion from the observation at the SK.

We did not consider the flavor conversion (the MSW effect)
in this paper although this can occur with appropriate
generation mixing and neutrino masses. With the same
$\Delta m^2 / E_\nu$, the resonance of the flavor conversion
occurs in more outer region than the resonance layer of the spin-flavor
conversion, because the matter potential for the flavor
conversion are not suppressed by $(1-2Y_e)$\cite{Athar-etal}.
Even if the flavor conversion occurs in more outer region,
the spectrum of $\bar\nu_e$ is not changed. There may be interesting
effect if we consider the conversion
in the iron core, or the mutual effect of spin-flavor and flavor
conversion among the three generations of neutrinos, as pointed out
by the earlier publication\cite{Athar-etal}.

The turbulence in the radial dependence of the magnetic fields
was ignored in this paper. In the solar metallicity models,
$\Delta H$ is much larger than $\mu_\nu B$ except at the
resonance, and the conversion probability is determined only from the
strength of the field at the resonance layer. Therefore the turbulence
does not affect the evolution of conversion probability of neutrinos.
However, the turbulence disturbs the relation
of $B_0$ and $B$ at the resonance
layer. When the neutrino energy changes, the location of the 
resonance layer also changes, and hence the neutrino spectrum 
can be disturbed by some strong turbulence in the magnetic fields.

The effect of dynamics in the collapse-driven supernova
was also not taken into consideration. Although it seems unlikely
that the shock wave is propagated 
through the whole isotopically neutral region
in a few tens of seconds, the inner part of the isotopically
neutral region may be dynamically disturbed by the shock wave. 
If $\Delta m^2 / E_\nu$
is large, the resonance occurs at the inner part of the isotopically
neutral region, and the dynamical effect may change the situation of
the resonant conversion of the neutrinos emitted in the later phase
of emission ($\agt$ 10 sec after the bounce). 
We give here simple discussion on the effect of the change in density,
assuming that $Y_e$ is conserved during the shock propagation.
(The composition of matter is drastically changed
by the shock wave, but the change in $Y_e$ requires the weak
interaction.) The matter potential ($\Delta V$) 
changes as $\propto \rho$. On the other hand,
if we assume the conservation of the magnetic flux in dynamical plasma,
the field strength changes as $\propto \rho^{2/3}$, and hence the precession
becomes more effective with decreasing matter density (see Eq.
(\ref{eq:precession-amplitude})). 
However, the adiabaticity condition is the competition of $\mu_\nu B$ and 
$|d(\Delta V)/dr|^{1/2}$, and if we assume 
$|d(\Delta V)/dr|$ scales as (length)$^{-4}$
(homogeneous expansion or compression),
the scaling of $|d(\Delta V)/dr|^{1/2}$ is the same with that of magnetic
fields, $\propto \rho^{2/3}$. Therefore, the adiabaticity condition
is not strongly affected by the dynamics of the shock wave.

It should also be noted that the observed data of neutrinos from
SN1987A \cite{SN1987A-Kam,SN1987A-IMB} favor a softer neutrino
spectrum than theoretically plausible spectrum of electron
antineutrinos. If $\bar\nu_e$'s are exchanged with $\nu_\mu$-like
neutrinos ($\nu_\mu$, $\nu_\tau$, and their antiparticles) which
have higher average energy, this discrepancy becomes larger.
From this view point, an upper bound on the conversion probability
of $\bar\nu_e$'s and $\nu_\mu$-like neutrinos has been derived: 
$P < 0.35$ at the 99 \% confidence level \cite{Smirnov-Spergel-%
Bahcall}. The earlier paper of the spin-flavor conversion 
\cite{Athar-etal} used this constraint in order to
derive an upper bound on the neutrino magnetic moment.
However, the above constraint on $P$ and its confidence level
suffer considerable statistical uncertainty because of the small number 
of events in Kamiokande and IMB. Therefore, we avoid
a decisive conclusion about the upper bound on $\mu_\nu$ from SN1987A,
although the strong conversion region ($P > 0.9$) in the contour maps
(Figs. \ref{fig:WW15S-contour}--\ref{fig:WW25Z-contour}) may be disfavored. 
Also we point out here that the metallicity effect on the conversion
probability should be taken into account when one attempts to constrain
$\mu_\nu$ from the SN1987A data, because the Large Magellanic 
Cloud is a low-metal system.

\section{Conclusions}
\label{sec:summary}
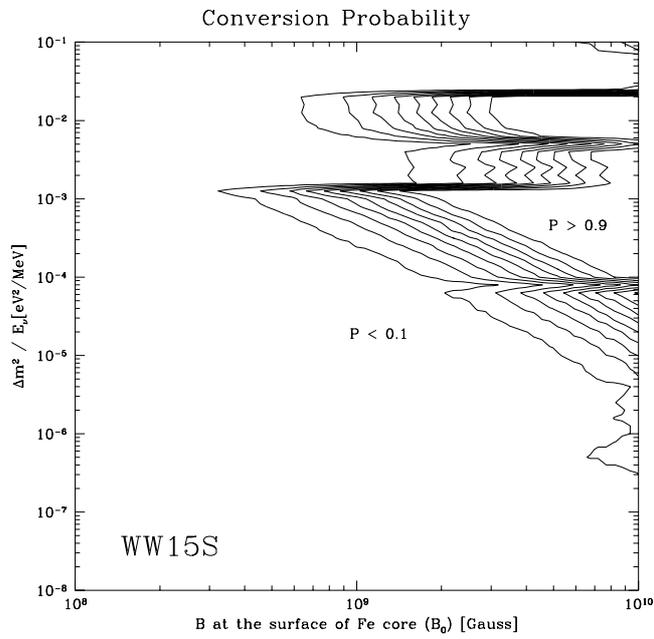
\begin{figure}
\caption{The contour map of conversion probability for the 
Woosley \& Weaver's solar-metallicity 15 $M_\odot$ model, 
as a function of 
$\Delta m^2 / E_\nu$ and $B_0$, where $B_0$ is the magnetic field 
at the surface of the iron core.
The neutrino
magnetic moment, $\mu_\nu$, is assumed to be $10^{-12}
\mu_B$. The contours are depicted with the interval of 0.1.
}
\label{fig:WW15S-contour}
\end{figure}

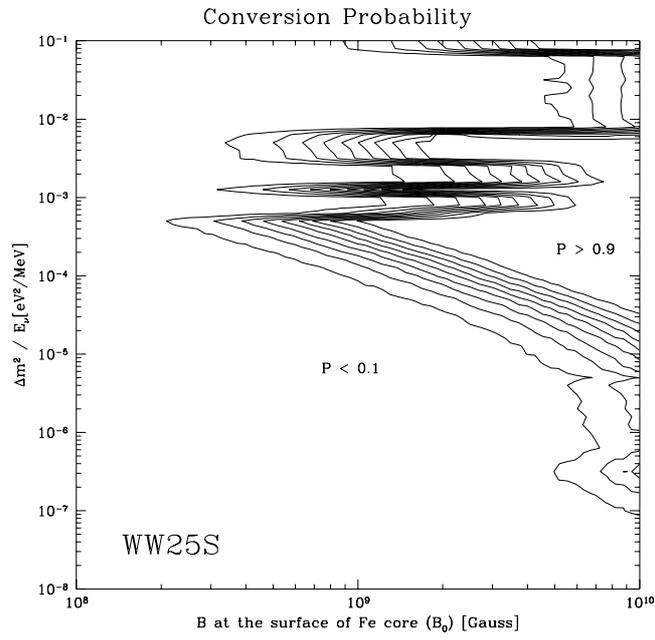
\begin{figure}
\caption{The same as Fig. \protect\ref{fig:WW15S-contour},
but for the Woosley \& Weaver's solar-metallicity 25 $M_\odot$ model.}
\label{fig:WW25S-contour}
\end{figure}

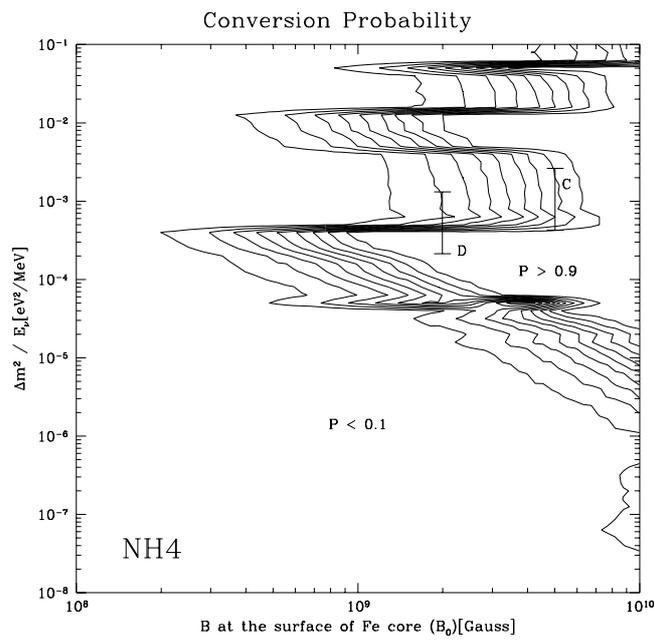
\begin{figure}
\caption{The same as Fig. \protect\ref{fig:WW15S-contour},
but for the Nomoto \& Hashimoto's 4 $M_\odot$ helium-core model.
The two vertical bars, (C) and (D) correspond to the energy
range of 10--70 MeV with 
($\Delta m^2$ [eV$^2$], $B_0$[Gauss]) = 
($3 \times 10^{-2}$, $5 \times 10^9$) and ($1.5 \times 10^{-2}$,
$2 \times 10^9$),  respectively,
where the energy range is that of positrons
which are observed in a water \v{C}erenkov detector.}
\label{fig:NH4-contour}
\end{figure}

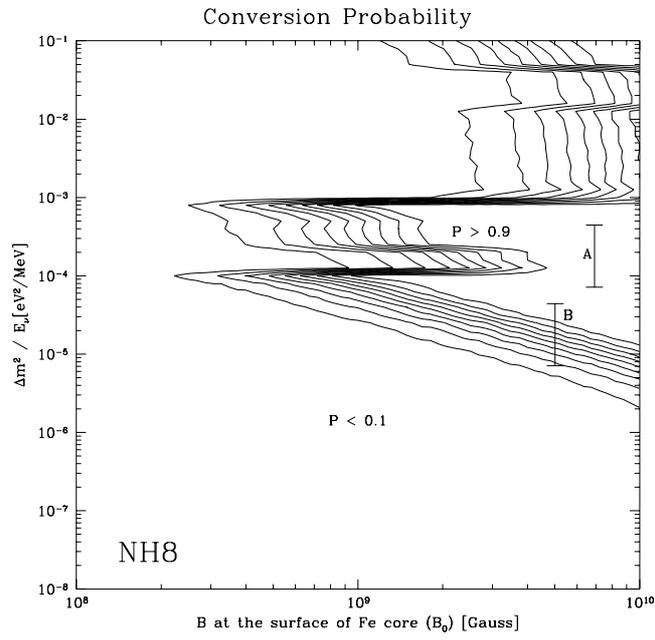
\begin{figure}
\caption{
The same as Fig. \protect\ref{fig:NH4-contour}, 
but for the Nomoto \& Hashimoto's 8 $M_\odot$ 
helium-core model. The values of ($\Delta m^2$
[eV$^2$], $B_0$ [Gauss]) for 
the two vertical bars, (A) and (B) are
($5 \times 10^{-3}$, $7 \times 10^9$) and ($5 \times 10^{-4}$,
$5 \times 10^9$), respectively.}
\label{fig:NH8-contour}
\end{figure}

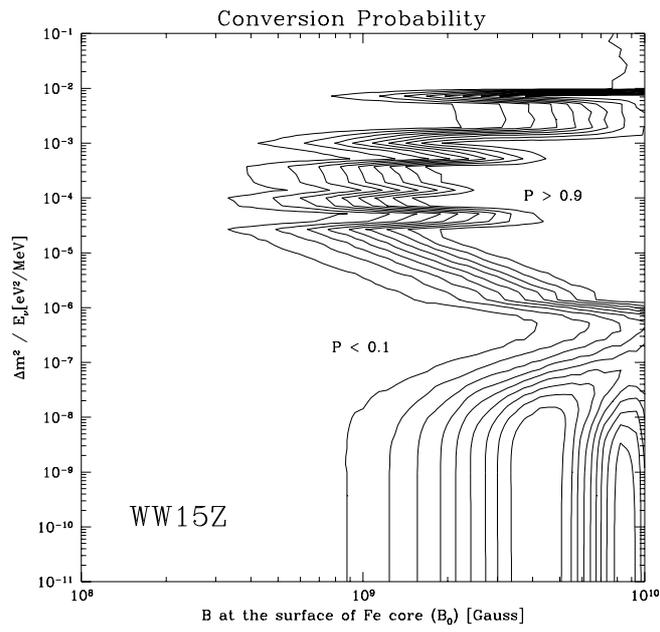
\begin{figure}
\caption{The same as Fig. \protect\ref{fig:WW15S-contour},
but for the Woosley \& Weaver's zero-metallicity 15 $M_\odot$ model. 
Note that the region of $\Delta m^2 / E_\nu$
is different from that in
Figs. \protect\ref{fig:WW15S-contour}--\protect\ref{fig:NH8-contour}.
}
\label{fig:WW15Z-contour}
\end{figure}

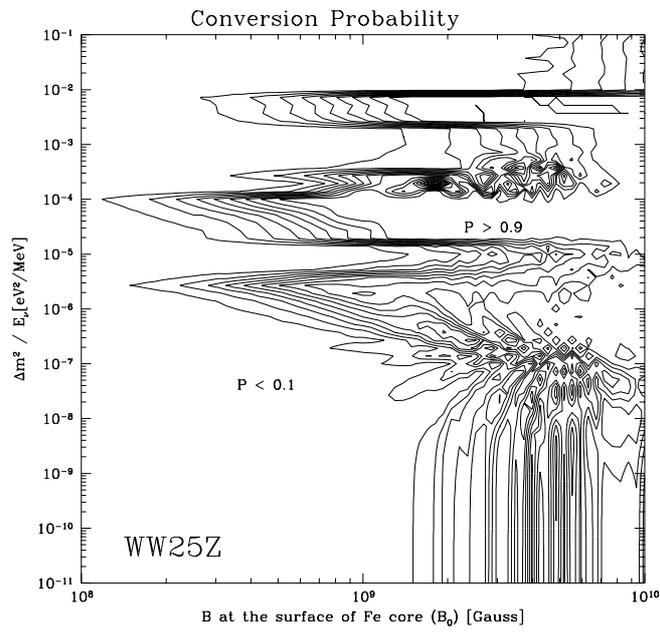
\begin{figure}
\caption{The same as Figure \protect\ref{fig:WW15Z-contour},
but for the Woosley \& Weaver's zero-metallicity 25 $M_\odot$ model. 
}
\label{fig:WW25Z-contour}
\end{figure}

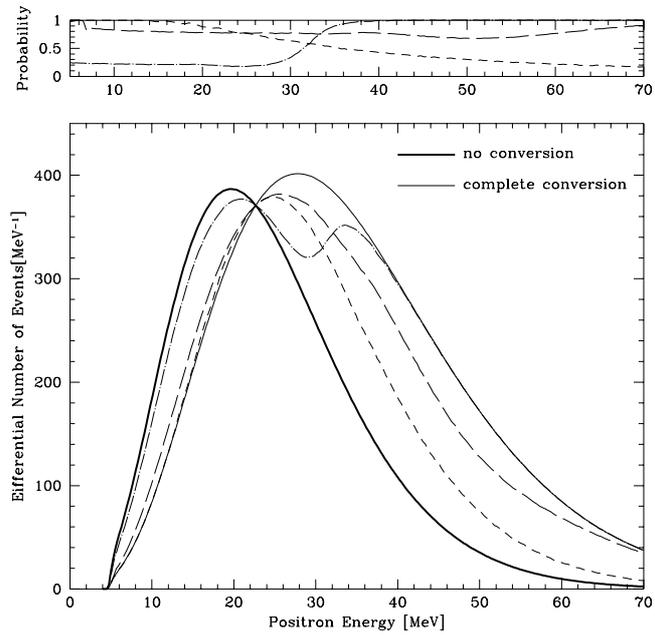
\begin{figure}
\caption{
Examples of the spectral deformation of $\bar\nu_e$'s
for the solar metallicity models.
The expected time-integrated differential
events at the SK detector are shown, assuming that the supernova
occurs in the Galactic center ($D = 10$ kpc). 
The upper box shows the corresponding
conversion 
probability of $\bar\nu_e \leftrightarrow \nu_\mu$.
The thick solid line shows the ordinary expected events of 
$\bar\nu_e$'s without any oscillation or conversion, and the thin solid line
the differential events when all of the original $\nu_\mu$'s are completely
converted to $\bar\nu_e$'s in the whole energy range. The short dashed line 
is the events when ($\Delta m^2$[eV$^2$], $B_0$[Gauss]) = 
(5 $\times 10^{-4}$, $5 \times 10^9$), using 
the model NH8, corresponding to the case (B) in Fig. \protect%
\ref{fig:NH8-contour}.
The long-dashed and dot-dashed lines are differential events 
with ($\Delta m^2$, $B_0$) = ($3 \times 10^{-2}$, $5 \times
10^9$) and ($1.5 \times 10^{-2}$, $2 \times 10^9$), respectively, 
and the model NH4 is used for both the lines.
The former corresponds to the bar (C) and the latter to the bar
(D) in Fig. \protect\ref{fig:NH4-contour}.}
\label{fig:spec-def-1}
\end{figure}

Neutrino spin-flavor conversion of $\bar\nu_e \leftrightarrow
\nu_\mu$ induced by the interaction of a flavor changing magnetic 
moment of Majorana neutrinos and magnetic fields 
above the iron core of collapsing stars was investigated
in detail. The effective matter potential of this conversion mode 
($\bar\nu_e \leftrightarrow \nu_\mu$)
is proportional to $(1-2Y_e)$, and hence this value is quite 
important to this resonant conversion. However, this 
value is determined by isotopes which are quite rarely existent,
and in order to estimate the effect of the astrophysical uncertainties,
we used the six precollapse models, changing
the stellar masses, metallicities, and authors of the models.
The components of Hamiltonian in the propagation 
equation (\ref{eq:schrodinger}) are shown for all the models
in Figs. \ref{fig:WW15S-ham}--\ref{fig:WW25Z-ham}, and 
qualitative features of the conversion can be understood from 
these figures. The results of the numerical calculation for all the 
models are shown in Figs. \ref{fig:WW15S-contour}--\ref{fig:%
WW25Z-contour} as contour maps of conversion probability
as a function of the two parameters of 
$\Delta m^2 / E_\nu$ and $B_0$, where $B_0$ is $B$ at the surface
of the iron core. 

For the solar metallicity models,
observable effects are expected when $\Delta m^2 / E_\nu$
is in the range of 
$10^{-5}$--$10^{-1}$ [eV$^2$/MeV] and $\mu_\nu \agt 
10^{-12} (10^9{\rm G}/B_0)$ [$\mu_B$]
(Figs. \ref{fig:WW15S-contour}--\ref{fig:NH8-contour}).
The difference of the stellar masses leads to the different 
thickness and location of the layers of the onion-like structure in
massive stars 
and this effect appears in the contour maps,
although the effect is rather small.
The qualitative features of the contour maps
for the WW and NH models are also not so different, and the model
dependence of the conversion probability can be roughly estimated
by the comparison of these figures.
Although the dependence on the stellar models or stellar masses
is rather weak as shown in Figs. 
\ref{fig:WW15S-contour}--\ref{fig:NH8-contour}, it was found that
the metal abundance of the precollapse star significantly affect the 
value of $(1-2Y_e)$. The difference between the solar and 
zero metallicity is prominent especially in the O+C and He layers,
and the strong precession between $\bar\nu_e \leftrightarrow \nu_\mu$
occurs with small $\Delta m^2 / E_\nu$,
because $\mu_\nu B$ is much larger than $\Delta m^2 / E_\nu$
in this region. In contrast to the solar metallicity models,
the conversion occurs even when $\Delta m^2 = 0$. The probability 
changes periodically
with $B_0$ because of the precession effect. (See Figs. 
\ref{fig:WW15Z-contour} and \ref{fig:WW25Z-contour}.)

Considering the above properties, the expected spectral
deformation of $\bar\nu_e$'s can be summarized as follows.
For the solar metallicity models,
there are roughly three types of the energy dependence of the 
conversion probability: 1) complete conversion
in a range of neutrino energy, 2) 
conversion probability decreases with increasing energy when
the energy range is in `the continuous deformation region' 
in the contour maps, and 3) incomplete conversion and weak
energy dependence of conversion probability when the energy range
is in `the weak adiabaticity band' in the contour maps. Examples of 
these types of spectral deformation are given in Fig. \ref{fig:spec-def-1}.
(For the explanation of `the continuous deformation region'
and `the weak adiabaticity band', see section \ref{sec:results}.)
Furthermore, there appear some interesting jumps in the spectrum if 
the energy range of 10--70 MeV includes
boundaries of the complete conversion region and 
the weak adiabaticity band. Irrespective of the stellar models or
stellar masses, such a boundary exists especially
at the surface of the iron core, where the matter potential suddenly
changes. For the zero metallicity models, although the feature 
of the spectral deformation is
similar to that of the solar metallicity models when
$\Delta m^2 / E_\nu \agt 10^{-7}$ [eV$^2$/MeV],
energy-independent conversion is possible with quite small 
$\Delta m^2 / E_\nu$ ($\alt 10^{-9}$ [eV$^2$/MeV]), which does not
occur in the solar metallicity models.

\section*{acknowledgments}
The authors would like to thank S.E. Woosley and M. Hashimoto,
for providing us the data of their precollapse models and useful comments.
They are also grateful to Y. Totsuka, for the information on
the detection efficiency of the SK detector. This work has been 
supported in part by the Grant-in-Aid for COE Research (07CE2002) and
for Scientific Research Fund (05243103 and 07640386) of the Ministry
of Education, Science, and Culture in Japan.

\vspace{2cm}
\begin{center}
\bf TABLES
\end{center}

\begin{table}
\caption{The table of $(1-2Y_e)$ in the isotopically neutral
region (O+Si, O+Ne+Mg, O+C, and He layers)
for all the models of precollapse stars used in this paper.
The models WW15 and WW25 are 15 and 25 $M_\odot$
models by Woosley and Weaver (1995) \protect%
\cite{WW1995}. There are models with
two different metallicities in the WW models,
and they are distinguished by `S' (solar metallicity) and `Z' 
(zero metallicity).
The models NH4 and NH8 are 4 and 8 $M_\odot$ helium core models
(corresponding approximately to 15 and 25 $M_\odot$ main sequence stellar
masses) by Nomoto and Hashimoto (1988)\protect\cite{NH1988}. The isotopes
in the parentheses under the values of $(1-2Y_e)$ are the main isotopes
which cause the deviation of $Y_e$ from 0.5.}
\label{table:ye}
\begin{tabular}{cdddd}
Model & O+Si & O+Ne+Mg & O+C & He \\
\hline
WW15S & 2 $\times 10^{-3}$   & 2 $\times 10^{-3}$ 
      & 2 $\times 10^{-3}$   & 1--10 $\times 10^{-4}$ \\

   &($^{34}$S, $^{38}$Ar)  & ($^{23}$Na, $^{25,26}$Mg) 
   &($^{22}$Ne, $^{25}$Mg) & ($^{18}$O, $^{56}$Fe) \\

WW25S & 2 $\times 10^{-3}$   & 2 $\times 10^{-3}$
      & 2 $\times 10^{-3}$   & 2 $\times 10^{-3}$  \\

   &($^{34}$S, $^{38}$Ar)     & ($^{25,26}$Mg,$^{27}$Al) 
   &($^{22}$Ne, $^{25,26}$Mg) & ($^{22}$Ne) \\

WW15Z & 6--10 $\times 10^{-4}$ & 4 $\times 10^{-4}$ 
      & 1--6 $\times 10^{-8}$   & 2--100 $\times 10^{-10}$ \\

   &($^{27}$Al, $^{29,30}$Si) &($^{23}$Na, $^{26}$Mg)
   &($^{22}$Ne, $^{26}$Mg)    &($^{18}$O) \\

WW25Z & 4 $\times 10^{-4}$   & 1 $\times 10^{-4}$
      & 2 $\times 10^{-7}$   & 2 $\times 10^{-7}$  \\

 &($^{27}$Al, $^{30}$Si) &($^{23}$Na, $^{26}$Mg) 
 &($^{22}$Ne, $^{25,26}$Mg)        &($^{13,14}$C, $^{22}$Ne) \\

NH4   & 1 $\times 10^{-3}$   & 2 $\times 10^{-3}$
      & 2 $\times 10^{-3}$   & 2 $\times 10^{-3}$  \\

 &($^{34}$S, $^{38}$Ar)  & ($^{25,26}$Mg, $^{27}$Al) 
 &($^{22}$Ne, $^{26}$Mg) & ($^{18}$O, $^{22}$Ne) \\

NH8   & 2 $\times 10^{-3}$   & 1 $\times 10^{-3}$
      & 1 $\times 10^{-3}$   & 2 $\times 10^{-3}$  \\

 &($^{27}$Al, $^{29,30}$Si) & ($^{23}$Na, $^{25,26}$Mg) 
 &($^{22}$Ne, $^{26}$Mg)    & ($^{18}$O, $^{22}$Ne) \\

\end{tabular}
\end{table}

\newpage
{\large \bf Figure Captions}\\

\end{document}